\documentclass[aps,prl,superscriptaddress,showpacs,floatfix,twocolumn]{revtex4}
\usepackage{graphicx}

\begin{document}

\title{System Size and Energy Dependence of Jet-Induced Hadron Pair
Correlation Shapes in Cu+Cu and Au+Au Collisions at 
$\sqrt{s_{NN}}$ = 200 and 62.4 GeV.}

\newcommand{\abilene}{Abilene Christian University, Abilene, TX 79699, U.S.}
\newcommand{\acadsin}{Institute of Physics, Academia Sinica, Taipei 11529, Taiwan}
\newcommand{\banaras}{Department of Physics, Banaras Hindu University, Varanasi 221005, India}
\newcommand{\barc}{Bhabha Atomic Research Centre, Bombay 400 085, India}
\newcommand{\bnl}{Brookhaven National Laboratory, Upton, NY 11973-5000, U.S.}
\newcommand{\caucr}{University of California - Riverside, Riverside, CA 92521, U.S.}
\newcommand{\charlesczech}{Charles University, Ovocn\'{y} trh 5, Praha 1, 116 36, Prague, Czech Republic}
\newcommand{\ciae}{China Institute of Atomic Energy (CIAE), Beijing, People's Republic of China}
\newcommand{\cns}{Center for Nuclear Study, Graduate School of Science, University of Tokyo, 7-3-1 Hongo, Bunkyo, Tokyo 113-0033, Japan}
\newcommand{\colorado}{University of Colorado, Boulder, CO 80309, U.S.}
\newcommand{\columbia}{Columbia University, New York, NY 10027 and Nevis Laboratories, Irvington, NY 10533, U.S.}
\newcommand{\czechtech}{Czech Technical University, Zikova 4, 166 36 Prague 6, Czech Republic}
\newcommand{\dapnia}{Dapnia, CEA Saclay, F-91191, Gif-sur-Yvette, France}
\newcommand{\debrecen}{Debrecen University, H-4010 Debrecen, Egyetem t{\'e}r 1, Hungary}
\newcommand{\elte}{ELTE, E{\"o}tv{\"o}s Lor{\'a}nd University, H - 1117 Budapest, P{\'a}zm{\'a}ny P. s. 1/A, Hungary}
\newcommand{\fit}{Florida Institute of Technology, Melbourne, FL 32901, U.S.}
\newcommand{\fsu}{Florida State University, Tallahassee, FL 32306, U.S.}
\newcommand{\gsu}{Georgia State University, Atlanta, GA 30303, U.S.}
\newcommand{\hiroshima}{Hiroshima University, Kagamiyama, Higashi-Hiroshima 739-8526, Japan}
\newcommand{\ihepprot}{IHEP Protvino, State Research Center of Russian Federation, Institute for High Energy Physics, Protvino, 142281, Russia}
\newcommand{\illuiuc}{University of Illinois at Urbana-Champaign, Urbana, IL 61801, U.S.}
\newcommand{\instpasczech}{Institute of Physics, Academy of Sciences of the Czech Republic, Na Slovance 2, 182 21 Prague 8, Czech Republic}
\newcommand{\isu}{Iowa State University, Ames, IA 50011, U.S.}
\newcommand{\jinrdubna}{Joint Institute for Nuclear Research, 141980 Dubna, Moscow Region, Russia}
\newcommand{\kaeri}{KAERI, Cyclotron Application Laboratory, Seoul, South Korea}
\newcommand{\kek}{KEK, High Energy Accelerator Research Organization, Tsukuba, Ibaraki 305-0801, Japan}
\newcommand{\kfki}{KFKI Research Institute for Particle and Nuclear Physics of the Hungarian Academy of Sciences (MTA KFKI RMKI), H-1525 Budapest 114, POBox 49, Budapest, Hungary}
\newcommand{\korea}{Korea University, Seoul, 136-701, Korea}
\newcommand{\kurchatov}{Russian Research Center ``Kurchatov Institute", Moscow, Russia}
\newcommand{\kyoto}{Kyoto University, Kyoto 606-8502, Japan}
\newcommand{\labllr}{Laboratoire Leprince-Ringuet, Ecole Polytechnique, CNRS-IN2P3, Route de Saclay, F-91128, Palaiseau, France}
\newcommand{\lawllnl}{Lawrence Livermore National Laboratory, Livermore, CA 94550, U.S.}
\newcommand{\losalamos}{Los Alamos National Laboratory, Los Alamos, NM 87545, U.S.}
\newcommand{\lpc}{LPC, Universit{\'e} Blaise Pascal, CNRS-IN2P3, Clermont-Fd, 63177 Aubiere Cedex, France}
\newcommand{\lund}{Department of Physics, Lund University, Box 118, SE-221 00 Lund, Sweden}
\newcommand{\muenster}{Institut f\"ur Kernphysik, University of Muenster, D-48149 Muenster, Germany}
\newcommand{\myongji}{Myongji University, Yongin, Kyonggido 449-728, Korea}
\newcommand{\nagasaki}{Nagasaki Institute of Applied Science, Nagasaki-shi, Nagasaki 851-0193, Japan}
\newcommand{\newmex}{University of New Mexico, Albuquerque, NM 87131, U.S. }
\newcommand{\nmsu}{New Mexico State University, Las Cruces, NM 88003, U.S.}
\newcommand{\ornl}{Oak Ridge National Laboratory, Oak Ridge, TN 37831, U.S.}
\newcommand{\orsay}{IPN-Orsay, Universite Paris Sud, CNRS-IN2P3, BP1, F-91406, Orsay, France}
\newcommand{\peking}{Peking University, Beijing, People's Republic of China}
\newcommand{\pnpi}{PNPI, Petersburg Nuclear Physics Institute, Gatchina, Leningrad region, 188300, Russia}
\newcommand{\riken}{RIKEN, The Institute of Physical and Chemical Research, Wako, Saitama 351-0198, Japan}
\newcommand{\rikjrbrc}{RIKEN BNL Research Center, Brookhaven National Laboratory, Upton, NY 11973-5000, U.S.}
\newcommand{\rikkyo}{Physics Department, Rikkyo University, 3-34-1 Nishi-Ikebukuro, Toshima, Tokyo 171-8501, Japan}
\newcommand{\saispbstu}{Saint Petersburg State Polytechnic University, St. Petersburg, Russia}
\newcommand{\saopaulo}{Universidade de S{\~a}o Paulo, Instituto de F\'{\i}sica, Caixa Postal 66318, S{\~a}o Paulo CEP05315-970, Brazil}
\newcommand{\seoulnat}{System Electronics Laboratory, Seoul National University, Seoul, South Korea}
\newcommand{\stonybrkc}{Chemistry Department, Stony Brook University, Stony Brook, SUNY, NY 11794-3400, U.S.}
\newcommand{\stonycrkp}{Department of Physics and Astronomy, Stony Brook University, SUNY, Stony Brook, NY 11794, U.S.}
\newcommand{\subatech}{SUBATECH (Ecole des Mines de Nantes, CNRS-IN2P3, Universit{\'e} de Nantes) BP 20722 - 44307, Nantes, France}
\newcommand{\tenn}{University of Tennessee, Knoxville, TN 37996, U.S.}
\newcommand{\titech}{Department of Physics, Tokyo Institute of Technology, Oh-okayama, Meguro, Tokyo 152-8551, Japan}
\newcommand{\tsukuba}{Institute of Physics, University of Tsukuba, Tsukuba, Ibaraki 305, Japan}
\newcommand{\vandy}{Vanderbilt University, Nashville, TN 37235, U.S.}
\newcommand{\waseda}{Waseda University, Advanced Research Institute for Science and Engineering, 17 Kikui-cho, Shinjuku-ku, Tokyo 162-0044, Japan}
\newcommand{\weizmann}{Weizmann Institute, Rehovot 76100, Israel}
\newcommand{\yonsei}{Yonsei University, IPAP, Seoul 120-749, Korea}
\affiliation{\abilene}
\affiliation{\acadsin}
\affiliation{\banaras}
\affiliation{\barc}
\affiliation{\bnl}
\affiliation{\caucr}
\affiliation{\charlesczech}
\affiliation{\ciae}
\affiliation{\cns}
\affiliation{\colorado}
\affiliation{\columbia}
\affiliation{\czechtech}
\affiliation{\dapnia}
\affiliation{\debrecen}
\affiliation{\elte}
\affiliation{\fit}
\affiliation{\fsu}
\affiliation{\gsu}
\affiliation{\hiroshima}
\affiliation{\ihepprot}
\affiliation{\illuiuc}
\affiliation{\instpasczech}
\affiliation{\isu}
\affiliation{\jinrdubna}
\affiliation{\kaeri}
\affiliation{\kek}
\affiliation{\kfki}
\affiliation{\korea}
\affiliation{\kurchatov}
\affiliation{\kyoto}
\affiliation{\labllr}
\affiliation{\lawllnl}
\affiliation{\losalamos}
\affiliation{\lpc}
\affiliation{\lund}
\affiliation{\muenster}
\affiliation{\myongji}
\affiliation{\nagasaki}
\affiliation{\newmex}
\affiliation{\nmsu}
\affiliation{\ornl}
\affiliation{\orsay}
\affiliation{\peking}
\affiliation{\pnpi}
\affiliation{\riken}
\affiliation{\rikjrbrc}
\affiliation{\rikkyo}
\affiliation{\saispbstu}
\affiliation{\saopaulo}
\affiliation{\seoulnat}
\affiliation{\stonybrkc}
\affiliation{\stonycrkp}
\affiliation{\subatech}
\affiliation{\tenn}
\affiliation{\titech}
\affiliation{\tsukuba}
\affiliation{\vandy}
\affiliation{\waseda}
\affiliation{\weizmann}
\affiliation{\yonsei}
\author{A.~Adare}	\affiliation{\colorado}
\author{S.S.~Adler}	\affiliation{\bnl}
\author{S.~Afanasiev}	\affiliation{\jinrdubna}
\author{C.~Aidala}	\affiliation{\columbia}
\author{N.N.~Ajitanand}	\affiliation{\stonybrkc}
\author{Y.~Akiba}	\affiliation{\kek}  \affiliation{\riken}  \affiliation{\rikjrbrc}
\author{H.~Al-Bataineh}	\affiliation{\nmsu}
\author{J.~Alexander}	\affiliation{\stonybrkc}
\author{A.~Al-Jamel}	\affiliation{\nmsu}
\author{K.~Aoki}	\affiliation{\kyoto} \affiliation{\riken}
% ...
\collaboration{PHENIX Collaboration} \noaffiliation

\date{\today}

\begin{abstract}
   We present azimuthal angle correlations of intermediate
transverse momentum ($1-4$ GeV/c) hadrons from dijets in
Cu+Cu and Au+Au collisions at $\sqrt{s_{NN}}$ = 62.4 and 200
GeV. The away-side dijet induced azimuthal correlation is
broadened, non-Gaussian, and peaked away from $\Delta\phi=\pi$ in
central and semi-central collisions in all the systems. The
broadening and peak location are found to depend upon the number
of participants in the collision, but not on the collision energy
or beam nuclei. These results are consistent with sound or shock
wave models, but pose challenges to Cherenkov gluon radiation
models.
\end{abstract}

\pacs{25.75.Dw}

\maketitle

    Heavy ion collisions at the Relativistic Heavy Ion Collider
(RHIC) produce QCD matter at enormous energy density
\cite{WhitePaper}, exceeding that required for a phase transition
to partonic, rather than hadronic, matter. The produced matter
exhibits collective motion \cite{PHENIXflow} and is opaque to
scattered quarks and gluons. The opacity is observed via
suppression of high momentum hadrons and intermediate energy
dijets \cite{PHENIXRaa}, and provides clear evidence of large
energy loss by partons (quarks or gluons) traversing the medium. A
key question is how the hot, dense medium transports the deposited
energy.

As partons fragment into back-to-back jets of hadrons, angular
correlations of the hadrons are used to study medium effects upon
hard scattered parton pairs.  Hadron pairs 
from the same parton appear at $\Delta\phi\sim 0$ (the near-side),
while those with one hadron from each parton in the hard scattered
pair appear at $\Delta\phi\sim\pi$ (the away-side). For brevity,
we will refer to these dijet induced dihadron azimuthal
correlations with the abbreviation ''dijet correlations''.

    Of great interest are intermediate transverse momentum
($p_{\rm T}$) hadrons, as they can arise from intermediate energy jets
or involve partons from the medium \cite{MachCones,WakeWaves}.
Their correlations can provide information about energy loss
mechanisms, dissipation of the radiated energy in the medium, and
collective modes induced by the deposited energy. Theoretical
ideas include: Mach cones from density waves induced by supersonic
partons \cite{MachCones}, co-moving radiated gluons producing
''wakes'' in the medium \cite{WakeWaves}, ultrarelativistic partons
creating Cherenkov gluon radiation \cite{CherenkovCones}, and
medium-induced gluon radiation at large emission angles
\cite{salgado,Ivan}. They all imply significant modifications of
dijet correlations in the away-side, when the parton path
through the medium is long. In particular, some of these
theoretical models \cite{MachCones,CherenkovCones,salgado} imply a
transition from the peaked distribution at $\Delta\phi\sim\pi$
characteristic of p+p and p+A collisions to a distribution with a
peak away from $\Delta\phi\sim\pi$ in head-on Au+Au collisions.

    Low $p_{\rm T}$ ($\ge 0.15$ GeV/c) hadrons associated with high
$p_{\rm T}$ hadrons ($\ge 4$ GeV/c) have modified away-side dijet
correlations and softened $p_{\rm T}$ distributions relative to those in
p+p collisions, suggesting that at least some of the lost energy
is thermalized in the medium \cite{STARlowpt}. At intermediate
$p_{\rm T}$, a strong non-Gaussian shape modification of the dijet
away-side correlation \cite{PPG032} indicates the possible
existence of a local minimum at $\Delta\phi=\pi$. This Letter
introduces new parameters to quantify this shape modification and
reports their dependence on collision energy, system size,
transverse momentum and centrality measured by the PHENIX
experiment at RHIC.

    The minimum bias data were collected in the years 2005 (Cu+Cu at
$\sqrt{s_{NN}}$ = 200 and 62.4 GeV), 2004 (Au+Au at
$\sqrt{s_{NN}}$ = 200 and 62.4 GeV), and 2003 (d+Au at
$\sqrt{s_{NN}}$ = 200 GeV). Charged hadrons are tracked using the
Drift Chambers and Pad Chambers of the PHENIX central arm
spectrometers at midrapidity ($|\eta|<0.35$) in the same way as
described in \cite{PPG032}. The number of events in the Au+Au data
at $\sqrt{s_{NN}}$ = 200 GeV used here is about 30 times higher
than that in \cite{PPG032}. Collision centrality and the number of
participant nucleons ($N_{part}$) are determined using the
Beam-Beam Counters (BBCs) and Zero Degree Calorimeters
\cite{CentDet}.

    Relative azimuthal distributions $Y_{same}(\Delta\phi)$
between "trigger" hadrons with $2.5<p_{\rm T}<4$ GeV/c and "associated"
hadrons with $1<p_{\rm T}<2.5$ GeV/c are formed. We correct their shape
for the non-uniform azimuthal acceptance of the PHENIX central
arms by using the mixed event pairs $Y_{mixed}(\Delta\phi)$
\cite{PPG032} from the same data sample:
\begin{equation}
    C(\Delta\phi) \equiv \frac{Y_{same}(\Delta\phi)}{Y_{mixed}(\Delta\phi)}
    \times \frac{\int Y_{mixed}(\Delta\phi) d\Delta\phi}
    {\int Y_{same}(\Delta\phi) d\Delta\phi}
    \label{eqn:CFDef}
\end{equation}
    Extensive Monte-Carlo simulations were performed to ensure
that the true pair distribution shape is recovered through this
procedure.

    In Au+Au and Cu+Cu collisions, hadrons have an azimuthal
correlation with the reaction plane orientation $\Phi_{RP}$ which
is proportional to $1+2v_2cos(2(\phi-\Phi_{RP}))$. This generates
a significant correlated background to our dijet source
$J(\Delta\phi)$ of azimuthal correlations:
\begin{equation}
    C(\Delta\phi) = b_0 (1 + 2\langle v_2^{assoc} \rangle \langle v_2^{trigg} \rangle
    \cos(2\Delta\phi)) + J(\Delta\phi) \label{eqn:CFSources}
\end{equation}

    The charged hadron $\langle v_2\rangle$, where
''$\langle\rangle$'' signifies an event average, was measured
through a reaction plane analysis using the BBCs ($3<|\eta|<4$) as
in \cite{PPG032,v2v4Data}.

    Hadrons have also a much smaller fourth order azimuthal
correlation with the reaction plane orientation. Its effect was
studied with the Au+Au data at 200 GeV by including the
corresponding $2\langle v_4^{assoc} \rangle \langle v_4^{trigg}
\rangle \cos(4\Delta\phi)$ component in the background term of Eq.
\ref{eqn:CFSources}, where the $\langle v_4\rangle$ values have
also been measured through the reaction-plane analysis
\cite{v2v4Data}. No significant $v_4$ systematic effects on the
shape of the dijet correlations were found.

    The background subtraction generates point-by-point
($\Delta\phi$ dependent) systematic errors from $\langle
v_2^{assoc}\rangle\langle v_2^{trigg}\rangle$ uncertainty and an
overall ($\Delta\phi$ independent) systematic error from $b_0$
uncertainty. The sources of $\langle v_2^{assoc}\rangle\langle
v_2^{trigg}\rangle$ uncertainty are the $\langle v_2\rangle$
systematic error \cite{PPG032}, dominated by the reaction plane
resolution uncertainty, the $\langle v_2\rangle$ statistical
error, and the systematic error from the $\langle v_2^{assoc}
\cdot v_2^{trigg} \rangle\approx\langle v_2^{assoc} \rangle \cdot
\langle v_2^{trigg} \rangle$ factorization approximation made in
Eq. \ref{eqn:CFSources}. The latter is estimated to be at most 5\%
of the $\langle v_2\rangle$ product for the most central events.

\begin{figure}[thb]
\includegraphics[width=1.0\linewidth]{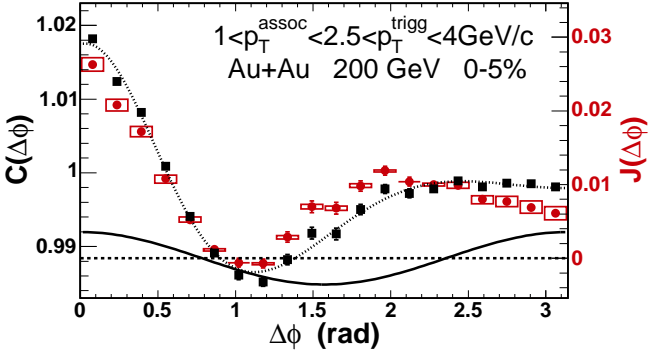}
\caption{\label{fig:JFExtract}
     (Color online) The measured correlation $C(\Delta\phi)$ (squares)
     and the dijet correlation $J(\Delta\phi)$ (circles
     with boxes for point-to-point systematic errors) in central
     Au+Au collisions at $\sqrt{s_{NN}} = 200$ GeV. The full line
     shows the background term and the dotted line shows a
     $C(\Delta\phi)$ fit with Eqs. (\ref{eqn:CFSources})+(\ref{eqn:JetFunc}).
     The left axis shows the measured correlation amplitude and
     the right axis shows the dijet correlation amplitude.}
\end{figure}

    The $b_0$ uncertainty is estimated by using three independent
methods to calculate $b_0$. The first, called Zero Yield At
Minimum (ZYAM), assumes that there is a region in $\Delta\phi$
where the dijet source of particle pairs is negligible. $b_0$ is
varied until the background component in Eq. \ref{eqn:CFSources}
matches the measured correlation $C(\Delta\phi)$ at some value of
$\Delta\phi$. In the second method a functional form for
$J(\Delta\phi)$ is added to the background, and the sum fitted to
the measured correlation with $b_0$ as a free parameter. Motivated
by the theoretical ideas discussed in the introduction, we use a
function that contains a near-side Gaussian, and two symmetric
away-side Gaussians:
\begin{equation} J(\Delta\phi) = G(\Delta\phi) + G(\Delta\phi-\pi-D)
+ G(\Delta\phi-\pi+D) \label{eqn:JetFunc}
\end{equation}
    While the choice of this functional form is not unique, it
does provide a reasonable fit to the measured correlations, as
shown by the dotted line in Fig. \ref{fig:JFExtract}. The
parameter D, or peak angle, is motivated by an attempt to describe
the away-side dijet correlation in terms of its symmetry around
$\Delta\phi\sim\pi$. We note that it also tends to absorb any
non-Gaussian character of the dijet correlation. The third
method is independent of the measured $C(\Delta\phi)$. We
calculate $b_{0} = \xi \kappa \langle n_{trigg}\rangle \langle
n_{assoc} \rangle /\langle n_{same} \rangle $ with hadron
production rates measured from all events within each centrality
class and scale by the same-event pair rate. $\kappa$ is a
correction for pair-cut bias and $\xi$ is a correction for
residual correlations due to averaging production rates from
events of different multiplicity within the same centrality class
\cite{MSMP}.

    As shown in Table \ref{tab:Errors} for the Au+Au data at 200
GeV, there are slight $b_0$ variations depending on which method
is used to extract its value. However, the resulting shape of the
dijet correlations is essentially independent of these
variations.

\begin{table}[thb]
\caption{\label{tab:Errors}
    $b_0$ values in Au+Au data at $\sqrt{s_{NN}} = 200$ GeV:
    ZYAM values (first row); variation of fit values from the
    ZYAM values (second row); variation of combinatorial values
    from the ZYAM values (third row).}
\begin{ruledtabular}
\begin{tabular}{ccccccc}
  Centrality  &60-90\% &40-60\% &20-40\% &10-20\% &5-10\% &0-5\%  \\ \hline
  ZYAM $b_0$  &0.861   &0.942   &0.960   &0.971   &0.982  &0.988  \\
  fit $\delta b_0$   &-0.003  &-0.003  &-0.006  &-0.028  &-0.035 &-0.022 \\
  comb. $\delta b_0$ &-0.086  &-0.013  &-0.004  &+0.002  &+0.001 &+0.001 \\
\end{tabular}
\end{ruledtabular}
\end{table}

    Figure \ref{fig:JFExtract} summarizes the extraction with the
ZYAM method of the dijet correlations using the central (0-5\%)
Au+Au data at $\sqrt{s_{NN}} = 200$ GeV: the measured correlation
is shown with squares, the background term with a full line, and
the background subtracted dijet correlation with circles for
values and boxes for the point-by-point systematic errors. The
systematic errors are correlated since they depend on the same
parameter - the $\langle v_2^{assoc}\rangle \langle
v_2^{trigg}\rangle$ uncertainty. For clarity, $J(\Delta\phi)$ is
shifted up by $b_0$, shown with dashed line, hence its amplitude
should be read from the right axis. We note that, in this case,
the measured correlation is flat near $\Delta\phi\sim\pi$, even
before any background subtraction. Due to the cosine modulation of
the background, a local minimum should develop at
$\Delta\phi\sim\pi$ in the dijet away-side correlation.

\begin{figure}[bht]
\includegraphics[width=1.0\linewidth]{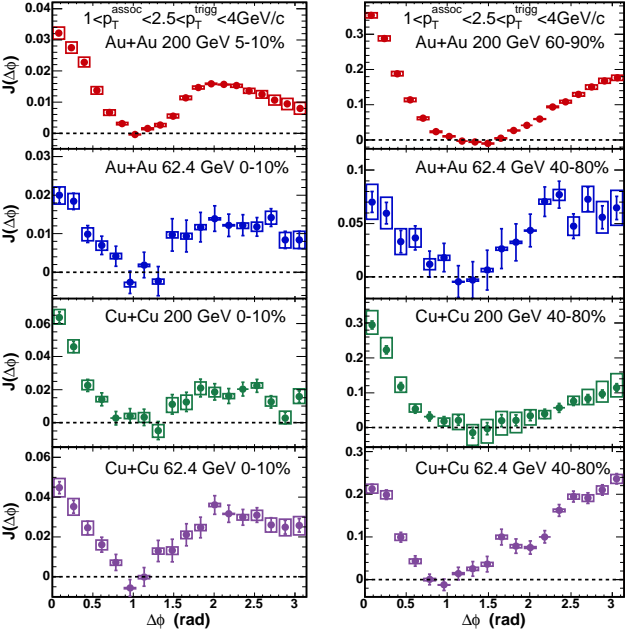}
\caption{\label{fig:AllJFs}
     (Color online) (Di)jet correlations (circles with boxes for
     point-to-point systematic errors) in Au+Au and Cu+Cu
     collisions at $\sqrt{s_{NN}}$ = 62.4 and 200 GeV.
     Left panels show central collisions, while right
     panels show peripheral collisions.}
\end{figure}

    Figure \ref{fig:AllJFs} shows a central and a peripheral
dijet correlation for each colliding system and energy. A
remarkable away-side feature in central and semi-central
collisions ($<40$\%) is the peak location away from
$\Delta\phi=\pi$, and the appearance of a local minimum at
$\Delta\phi=\pi$. To quantify the significance of this minimum in
the Au+Au data at 200 GeV, we have studied how much $\langle
v_2^{assoc} \rangle \langle v_2^{trig} \rangle$ would need to
change for the away-side to be flat. For the four most central
bins (0-5\%, 5-10\%, 10-20\%, and 20-40\%) it would have to
decrease by 85\%(5.1$\sigma$), 41\%(4.2$\sigma$),
20\%(2.3$\sigma$), and 23\%(2.7$\sigma$), respectively, where
$\sigma$ is the total $\langle v_2^{assoc} \rangle \langle
v_2^{trig} \rangle$ uncertainty.

    We parameterize the away-side shape change and deviation
from a Gaussian distribution by extracting the second and fourth
central moments around $\Delta\phi\sim\pi$ ($\mu_n\equiv\langle
(\Delta\phi-\pi)^n \rangle , n=2,4$), in the standard form of the
following statistical quantities: the root mean square rms
$\equiv\,\sqrt{\mu_2}$ and the kurtosis $\equiv\,\mu_4/\mu_2^2$.
The away-side is defined here as all $\Delta\phi$ values above the
dijet function $J(\Delta\phi)$ minimum, typically one rad. We
extract these statistics on only the away-side jet peaks in
$J(\Delta\phi)$; possible jet-associated flat underlying
distributions, which are highly sensitive to the uncertainty in
$b_{0}$ and precluded by the ZYAM assumption, are not included.

\begin{figure}[thb]
\includegraphics[width=1.0\linewidth]{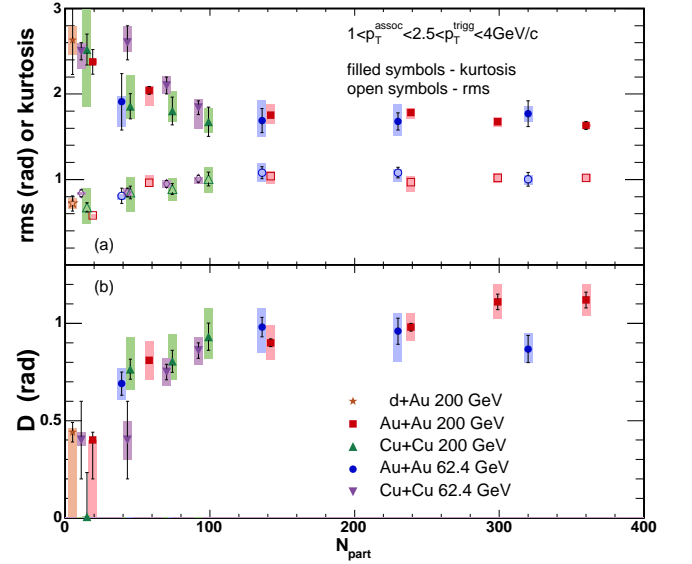}
\caption{\label{fig:ShapePars}
     (Color online) Collision centrality, energy, and system
     size dependence of shape parameters: (a) kurtosis (filled symbols)
     and rms (open symbols); (b) peak angle D. Bars show statistical
     errors, shaded bands systematic errors.}
\end{figure}

    The rms and kurtosis centrality dependence is shown in Fig.
\ref{fig:ShapePars}(a). The rms increases with centrality,
indicating broadening of the away-side dijet correlation, while
the kurtosis decreases from the value characteristic of a Gaussian
shape (three), suggesting a flattening of its shape beyond an
increase in the Gaussian width.

\begin{table}[bh]
\caption{\label{tab:MomDep} Dependence of away-side shape
parameters on associated hadron $p_{\rm T}$ in central (0-20\%) Au+Au
collisions at $\sqrt{s_{NN}} = 200$ GeV for $3<p_{\rm T}^{trigg}<5$
GeV/c. First error is statistical and second error is systematic.}
\begin{ruledtabular}
\begin{tabular}{cccc}
 $p_{\rm T}^{assoc}$ &D [rad]                &rms [rad]              &kurtosis  \\ \hline
 1-1.5 &1.04$\pm$0.03$\pm$0.03 &1.02$\pm$0.02$\pm$0.05 &1.68$\pm$0.04$\pm$0.10 \\
 1.5-2 &1.07$\pm$0.04$\pm$0.04 &1.06$\pm$0.02$\pm$0.05 &1.58$\pm$0.05$\pm$0.10 \\
 2-2.5 &1.05$\pm$0.03$\pm$0.06 &1.08$\pm$0.04$\pm$0.08 &1.38$\pm$0.11$\pm$0.12 \\
 2.5-3 &1.07$\pm$0.06$\pm$0.06 &1.09$\pm$0.07$\pm$0.07 &1.35$\pm$0.17$\pm$0.12 \\
 3-5   &0.88$\pm$0.13$\pm$0.16 &1.01$\pm$0.11$\pm$0.14 &1.31$\pm$0.23$\pm$0.25 \\
\end{tabular}
\end{ruledtabular}
\end{table}

    The peak angle D centrality dependence, extracted by
fitting dijet correlations with Eq. \ref{eqn:JetFunc}, is shown
in Fig. \ref{fig:ShapePars}(b). It is consistent with zero radians
in d+Au and peripheral nuclear collisions, but rapidly grows to a
value around one radian in central nuclear collisions. Some
deviation from zero radians of the peak angle may be related to
slight non-Gaussian shapes of the dijet correlations even
without medium modification. This can be seen in the kurtosis
values for d+Au and peripheral nuclear collisions, which have
values somewhat lower than three. For details on dijet
correlations in d+Au see \cite{dAuData}. The systematic errors in
Fig. \ref{fig:ShapePars} come exclusively from $v_2$ uncertainty.
No apparent dependence of rms, kurtosis, or peak angle D on
collision energy or species is observed.

    Table \ref{tab:MomDep} shows the dependence of the away-side
shape parameters on the associated hadron $p_{\rm T}$ in the Au+Au data
at 200 GeV for a 0-20\% centrality bin, $3<p_{\rm T}^{trigg}<5$ GeV/c,
and the following $p_{\rm T}^{assoc}$ bins: $1-1.5$, $1.5-2$, $2-2.5$,
$2.5-3$, and $3-5$ GeV/c. The peak angle D and the rms have no
$p_{\rm T}$ dependence, while the kurtosis is consistent with a slow
decrease with $p_{\rm T}$.

    Several phenomenological models for modification of the
away-side jet have been proposed; all involve a strong response of
the medium to the traversing jet. Bow shocks propagating as sound,
or density, waves in the medium produce a peak located away from
$\Delta\phi=\pi$ at approximately the same angle as seen in the
data \cite{MachCones,renkrup}. If the peak indeed arises from a
sound wave, its location at one radian away from the nominal jet
direction implies a speed of sound intermediate between that
expected in a hadron gas and quark-gluon plasma \cite{MachCones}.
A first order phase transition would cause a region with speed of
sound identically zero. This region was postulated
\cite{MachCones} to reflect sound waves and cause a second
away-side peak located at about 1.4 radians away from
$\Delta\phi=\pi$. No clear evidence for a distinct peak is seen in
our data.

    If the coupling among partons in the medium is strong, then 
the high momentum parton may induce non-sound wave collective 
plasma excitations \cite{WakeWaves}.  In the strong coupling 
limit the AdS/CFT correspondence was applied to calculate 
the wake of directional emission from a heavy quark traversing 
the medium, where a peak angle is found at values slightly 
larger than in these data \cite{gubser}.

    The peak may also arise from Cherenkov gluon radiation
\cite{CherenkovCones}. Such a mechanism should disappear for high
energy gluons, implying that the peak angle D should gradually
approach zero with increasing momentum of associated hadrons.
Table \ref{tab:MomDep} shows that this is not supported by the
data. The medium may induce gluon radiation at large angles by
mechanisms other than Cherenkov radiation \cite{salgado,Ivan}.
Such models can reproduce the observed peak if the density of
scattering centers is large and the gluon splitting sufficiently
asymmetric \cite{salgado}. However, the predicted radiation is
very sensitive to the treatment of geometry, expansion and
radiative energy loss framework used. Our detailed measurements
constrain the options.

    An important issue is whether the density wave correlations
can survive the underlying medium expansion. It was shown that the
interplay of the longitudinal expansion and limited experimental
$\eta$ acceptance preserves, and even amplifies, the signal of
directed collective excitations \cite{renkrup}. However, the
creation of a shock wave consistent with our data requires that
75-90\% of the jet's lost energy be transferred to the collective
mode \cite{renkrup,MachCones,chaudhuri}.

    We have presented azimuthal angle correlations of
intermediate transverse momentum hadrons from dijets in
Cu+Cu and Au+Au collisions at $\sqrt{s_{NN}}$ = 62.4 and 200 GeV.
The away-side dijet correlation is seen to be broadened,
non-Gaussian and peaked away from $\Delta\phi=\pi$ in central and
semi-central collisions. The away-side shape depends on the number
of participants in the collision, and not on the beam nuclei or
energy. The general features of the observed shape can be
qualitatively accounted for by a number of phenomenological
models, all having in common a strong medium response to the
energy deposited by the traversing parton. The systematic data
presented here provide quantitative tests that could discriminate
between these models.

\begin{acknowledgments}
    We thank the staff of the Collider-Accelerator and Physics
Departments at BNL for their vital contributions. We acknowledge
support from the Department of Energy and NSF (U.S.A.), MEXT and
JSPS (Japan), CNPq and FAPESP (Brazil), NSFC (China), MSMT (Czech
Republic), IN2P3/CNRS and CEA (France), BMBF, DAAD, and AvH
(Germany), OTKA (Hungary), DAE (India), ISF (Israel), KRF and
KOSEF (Korea), MES, RAS, and FAAE (Russia), VR and KAW (Sweden),
U.S. CRDF for the FSU, US-Hungarian NSF-OTKA-MTA, and US-Israel
BSF.
\end{acknowledgments}

\end{document}